\documentclass{emulateapj}
\usepackage{apjfonts}
\newcommand{\figexpand}{\epsscale{1.15}}
\newcommand{\tableclear}{}
\newcommand{\tableset}{deluxetable}
\newcommand{\plotter}{\plotone}

\newcommand{\etal}{et al.}
\newcommand{\mbh}{M_{\rm BH}}
\newcommand{\mstar}{M_{\ast}}
\newcommand{\mdyn}{M_{\rm dyn}}
\newcommand{\re}{R_{e}}

\shorttitle{A Black Hole Fundamental Plane}
\shortauthors{Hopkins \etal}
\slugcomment{Accepted to ApJ, May 2007}
\begin{document}

\title{An Observed Fundamental Plane Relation for Supermassive Black Holes}
\author{Philip F.\ Hopkins\altaffilmark{1}, 
Lars Hernquist\altaffilmark{1}, 
Thomas J. Cox\altaffilmark{1}, 
Brant Robertson\altaffilmark{2}, \&\
Elisabeth Krause\altaffilmark{3}
}
\altaffiltext{1}{Harvard-Smithsonian Center for Astrophysics, 60 Garden Street, Cambridge, MA 02138}
\altaffiltext{2}{Kavli Institute for Cosmological Physics, The University of Chicago, 
5460 S.\ Ellis Ave., Chicago, IL 60637}
\altaffiltext{3}{Department of Physics and Astronomy, Universit\"{a}t Bonn, 53121 Bonn, Germany}

\begin{abstract}

We study observed correlations between supermassive black hole (BHs)
and the properties of their host galaxies, and show that the
observations define a BH ``fundamental plane'' (BHFP), of the form
$\mbh\propto\sigma^{3.0\pm0.3}\,\re^{0.43\pm0.19}$ or
$\mbh\propto\mstar^{0.54\pm0.17}\,\sigma^{2.2\pm0.5}$, analogous to
the FP of elliptical galaxies. The BHFP is preferred over a simple
relation between $\mbh$ and any of $\sigma$, $\mstar$, $\mdyn$, or
$\re$ alone at $>3\,\sigma$ ($99.9\%$) significance. The existence of
this BHFP has important implications for the formation of supermassive
BHs and the masses of the very largest black holes, and immediately
resolves several apparent conflicts between the BH masses expected and
measured for outliers in both the $\mbh-\sigma$ and $\mbh-\mstar$
relations.

\end{abstract}

\keywords{quasars: general --- galaxies: active --- 
galaxies: evolution --- cosmology: theory}

\section{Introduction}
\label{sec:intro}

Discoveries of correlations between the masses of supermassive black
holes (BHs) in the centers of nearby galaxies and the properties of
their host spheroids \citep[e.g.,][]{KormendyRichstone95} demonstrate
a fundamental link between the growth of BHs and galaxy formation. A
large number of similar correlations have now been identified, linking
BH mass to host luminosity \citep{KormendyRichstone95}, mass
\citep{magorrian}, velocity dispersion \citep{FM00,Gebhardt00},
concentration or Sersic index
\citep{graham:concentration,graham:sersic}, and binding energy
\citep{aller:mbh.esph}, among others.  However, because these
properties of host spheroids are themselves correlated, it is not
clear whether any are in some sense more basic \citep[see e.g.][for
such a comparison]{novak:scatter}.

The lack of a clear motivation for favoring one relation over another
has led to substantial observational and theoretical debate over the
``proper'' correlation for systems which may not lie on the mean
correlation between host properties, and over the demographics of the
most massive BHs \citep[e.g.,][]{bernardi:magorrian.bias,
lauer:massive.bhs,batcheldor:bcgs,wyithe:log.quadratic.msigma}.  One
possibility is that these different correlations are projections of
the same ``fundamental plane'' (FP) relating BH mass with two or more
spheroid properties such as stellar mass, velocity dispersion, or
effective radius, in analogy to the well-established fundamental plane
of spheroids. For the case of spheroids, it is now understood that
various correlations, including the Faber-Jackson relation
\citep{fj76} between luminosity (or effectively stellar mass $\mstar$)
and velocity dispersion $\sigma$, the \citet{kormendy77:correlations}
relation between effective radius $\re$ and surface brightness
$I_{e}$, and the size-luminosity or size-mass relations
\citep[e.g.,][]{shen:size.mass} between $\re$ and $\mstar$, are all
projections of a fundamental plane relating $\re\propto
\sigma^{\alpha}\,I_{e}^{\beta}$ \citep{dressler87:fp,dd87:fp}.

\begin{figure}
    \centering
    \plotter{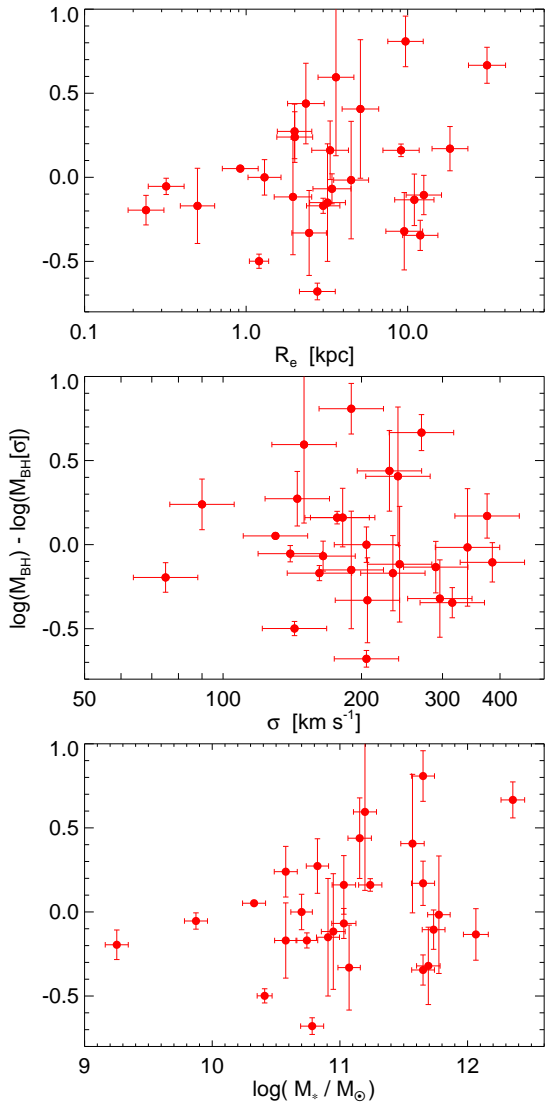}
    \caption{Observed residual in BH mass relative to the best-fit $\mbh-\sigma$ relation, 
    as a function of the host galaxy effective radius $\re$, velocity dispersion $\sigma$, 
    and stellar mass $\mstar$. There is a clear suggestion in 
    the data of a residual correlation 
    with $\re$ and $\mstar$, which we examine in detail in Figure~\ref{fig:data.residuals}.
    \label{fig:demo.residuals}}
\end{figure}

In their analysis of the relation between BH mass and host luminosity
or dynamical mass, $\mdyn$, \citet{marconihunt} \citep[see
also][]{defrancesco:mbh.mdyn} noted that the residuals of the
$\mbh-\sigma$ relation \citep[effectively
$\mbh/\sigma^{4}$;][]{tremaine:msigma} were significantly correlated
with the effective radii of the systems in their sample.
Figure~\ref{fig:demo.residuals} shows this, for the compilation of
observations described in \S~\ref{sec:data} -- there is indeed a clear
trend that systems with larger effective radii or stellar masses tend
to have larger $\mbh / \langle\mbh(\sigma)\rangle$.  In that context,
the authors argued for this as evidence favoring a relation between
$\mbh$ and $\mdyn\propto \sigma^{2}\,\re$ over $\mbh\propto
\sigma^{4}$, but it is not clear that a dependence on $\mdyn$ alone
completely or accurately captures the behavior in these residuals (and
in \S~\ref{sec:local:FP} we show that it does not). Furthermore,
finding $\mbh/\sigma^{4} \propto \re^{\beta}$ does not necessarily
imply a FP-like relation, if the correlation between $\mbh$ and
$\sigma$ or $\sigma$ and $\re$ has some nonlinear
\citep[e.g.,][]{wyithe:log.quadratic.msigma} or otherwise incompletely
accounted-for behavior. Still, this brings up the important
possibility of a true FP-like relation in which the combination of two
properties such as $\mstar$ and $\sigma$ determines $\mbh$, which we
study herein.

Throughout, we adopt a $\Omega_{\rm M}=0.3$, $\Omega_{\Lambda}=0.7$,
$H_{0}=70\,{\rm km\,s^{-1}\,Mpc^{-1}}$ cosmology 
(and correct all observations accordingly), but note this choice has little 
effect on our conclusions.

\section{The Data}
\label{sec:data}

We consider the sample of local BHs for which masses have been
reliably determined via either kinematic or maser
measurements. Specifically, we adopt the sample of 38 local systems
for which values of $\mbh$, $\sigma$, $\re$, $\mdyn$, and bulge
luminosities are compiled in \citet{marconihunt} and \citet{haringrix}
\citep[see also][]{magorrian,
merrittferrarese:msigma,tremaine:msigma}.  We adopt the dynamical
masses from the more detailed Jeans modeling in \citet{haringrix}. We
estimate the total stellar mass $\mstar$ from the total $K$-band
luminosity given in \citet{marconihunt}, using the $K$-band
mass-to-light ratios as a function of luminosity from \citet{bell:mfs}
(specifically assuming a ``diet'' Salpeter IMF, although this only
affects the absolute normalization of the relevant relations). We have 
repeated these calculations using mass-to-light ratios estimated 
independently for each object from their $B-J$ colors or 
full $UBVJHK$ photometry, and find it makes little difference. Where
possible, we update measurements of $\re$ and $\sigma$ with more
recent values from
\citet{lauer:centers,lauer:bimodal.profiles,mcdermid:sauron.profiles}
and from \citet{kormendy:wetvsdry}.

Although it should only affect the normalization of the relations
herein, we note that our adopted cosmology is identical to that used
to determine all quoted values in these works.  When we fit the
observations to e.g.\ the mean $\mbh-\sigma$ relation and other
BH-host relations, we consider only the subsample of 27 objects in
\citet{marconihunt} which are deemed to have 'secure' BH and bulge
measurements (i.e.\ for which the BH sphere of influence is clearly
resolved, the bulge profile can be well-measured, and maser spots
(where used to measure $\mbh$) are in Keplerian orbits).  Our results
are not qualitatively changed if we consider the entire sample in
these fits, but their statistical significance is somewhat reduced.

\section{A Black Hole Fundamental Plane}
\label{sec:local:FP}

We wish to determine whether or not a simple one-to-one correlation between e.g.\ 
$\mbh$ and $\sigma$ is a sufficient description of the data, or 
if there is evidence for additional dependence on a second parameter such as 
$\re$ or $\mstar$. The most efficient 
way to determine such a dependence is by looking for correlations 
between the 
residuals of the various projections of such a potential BHFP relation. 
In Figure~\ref{fig:demo.residuals}, we follow \citet{marconihunt} and 
plot the dependence on the 
residual of the $\mbh-\sigma$ relation on $\re$ and $\mstar$. This 
provides a clear suggestion of a residual dependence, and motivates 
us to examine a more complete description than a simple $\mbh-\sigma$ 
relation. However, in order to robustly do so, we must consider 
the correlations between residuals at fixed $\sigma$, not the correlation 
between e.g.\ the residual of $\mbh-\sigma$ and the actual value of 
a quantity such as $\re$ or $\mstar$. 

\begin{figure*}
    \centering
    \figexpand
    \plotone{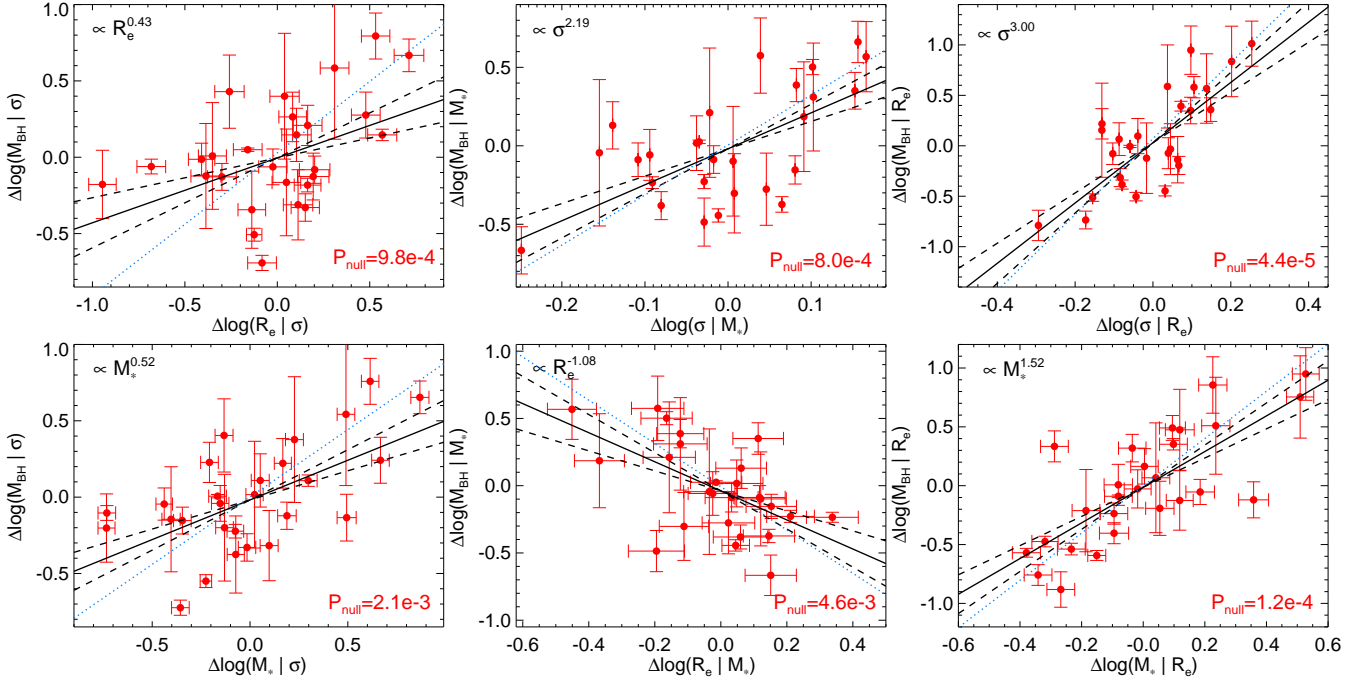}
    \caption{{\em Left:} Observed residual in BH mass as a function 
    of host galaxy effective radius $\re$ ({\em top}) or 
    stellar mass $\mstar$ ({\em bottom}), at fixed velocity dispersion $\sigma$ 
    (equivalently, correlation between the residuals in the $\mbh-\sigma$ 
    and $\re-\sigma$ or $\mstar-\sigma$ relations at each $\sigma$). 
    The fit to this residual correlation is 
    shown with the black lines ($\pm1\,\sigma$ range in the best-fit correlation shown 
    as dashed lines -- note that they are strongly inconsistent with zero correlation), 
    with the slope shown (dotted blue line shows the least-squares bisector). 
    The probability of the null hypothesis of no correlation in the 
    residuals (i.e.\ no systematic dependence of $\mbh$ on $\re$ or $\mstar$ at fixed $\sigma$) for 
    the observed systems is shown (red $P_{\rm null}$). 
    {\em Center:} Residual in $\mbh$ as a function of 
    $\sigma$ or $\re$ at fixed stellar mass $\mstar$. 
    {\em Right:} Residual in $\mbh$ as a function of $\sigma$ or $\mstar$ at 
    fixed effective radius $\re$. 
    The observations imply 
    a secondary ``fundamental plane''-type correlation at $3\,\sigma$ with respect to 
    each of these variables. 
    \label{fig:data.residuals}}
\end{figure*}

Figure~\ref{fig:data.residuals} plots the correlation between BH mass $\mbh$ and host 
bulge effective radius $\re$ or bulge stellar mass $\mstar$, 
all at fixed $\sigma$. Specifically, we determine the 
residual with respect to the $\mbh-\sigma$ relation by 
fitting $\mbh(\sigma)$ to an arbitrary log-polynomial 
\begin{equation}
\langle\log(\mbh)\rangle=\Sigma\,{\bigl[}\,a_{n}\,\log(\sigma)^{n}{\bigr]}, 
\end{equation}
allowing as many terms as the data favor (i.e.\ until $\Delta\chi^{2}$ with respect to 
the fitted relation is $<1$), and then taking 
\begin{equation}
\Delta\log(\mbh\,|\,\sigma)\equiv \log(\mbh) - \langle\log(\mbh)\rangle(\sigma).
\end{equation}
We determine the residual $\Delta\log(\re\,|\,\sigma)$
(or $\Delta\log(\mstar\,|\,\sigma)$, for the stellar mass) in identical fashion, and 
plot the correlation between the two. 
We allow arbitrarily high terms in $\log(\sigma)$ to avoid introducing bias 
by assuming e.g.\ a simple power-law correlation between $\mbh$ and $\sigma$, but 
find in practice that such terms are not needed -- as discussed below, there is 
no significant evidence for a log-quadratic (or higher-order) dependence 
of $\mbh$ on $\sigma$, $\re$, or $\mstar$, so allowing for these terms 
changes the residual best-fit solutions in Figure~\ref{fig:data.residuals} 
by $\ll 1\sigma$. For reference, we find (in this sample) a best-fit correlation 
between $\re\propto\sigma^{2.0\pm0.4}$ (ignoring the log-quadratic and higher order 
terms, which are not significant), similar to that found for all SDSS galaxies 
by \citet{shen:size.mass}, and $\mstar\propto\sigma^{3.4\pm0.5}$, again similar to 
the Faber-Jackson relation estimated for global populations \citep[e.g.][]{bernardi:correlations}. 
In any case, changing these slopes 
slightly (or e.g.\ using the $\mbh-\sigma$ relation of \citet{FM00}, 
as opposed to our best-fit which is very similar to \citet{tremaine:msigma}) has a 
relatively small systematic 
effect on the best-fit residual slopes (this generally makes the trend in residuals 
stronger, because the ``mean'' relation being subtracted is slightly less accurate 
for this particular sample), but does not qualitatively change our conclusions. 

Of course, even this approach could 
in principle introduce a bias via our assumption of some functional form, and 
so we have also considered a non-parametric approach where 
we take the mean $\langle \log(\mbh) \rangle$ in bins of $\log(\sigma)$. 
Unfortunately, 
the small number of observations limits such a non-parametric technique and 
somewhat smears out the interesting correlations (slightly decreasing their statistical significance). 
Regardless, the lack of evidence for a non-linear (i.e.\ log-quadratic or 
higher-order) relation means that we find similar results in 
all these cases, so we conclude that our methodology in determining residuals 
is not introducing a significant bias.

The figure demonstrates that there is a highly significant correlation between 
$\mbh$ and $\re$ or $\mstar$ at fixed $\sigma$. We repeat this exercise in the 
figure, and demonstrate similarly that there is a highly significant correlation 
between $\mbh$ and $\sigma$ or $\mstar$ at fixed $\re$, and 
between $\mbh$ and $\sigma$ or $\re$ at fixed $\mstar$. This indicates 
that a simple one-variable correlation (e.g.\ a 
$\mbh(\sigma)$, or $\mbh(\mstar)$, $\mbh(\re)$ relation) is an 
incomplete description of the observations. We therefore 
introduce a FP-like relation of the form 
\begin{equation}
\mbh\propto\sigma^{\alpha}\,\re^{\beta} \, ,
\end{equation}
which can account for these dependencies. 
Formally, we determine the combination of $(\alpha,\,\beta)$ which simultaneously 
minimizes the $\chi^{2}/\nu$ of the fit and 
the significance of the correlations between the residuals in $\sigma$ and 
$\mbh$ (or $\re$ and $\mbh$). This yields similar results to 
the direct fitting method of 
\citet{bernardi:fp} from the spheroid FP, which minimizes 
\begin{equation}
\Delta^{2} = {\bigl[} \log(\mbh) - \alpha\log(\sigma) - \beta\log(\re) - \delta {\bigr]}^{2}.
\end{equation}
It is straightforward to extend this minimization by weighting 
each point by the measurement errors (where we allow for the 
errors in all observed quantities -- $\log(\mbh)$, $\log(\sigma)$, 
and $\log(\re)$, and estimate symmetric errors as the mean of 
quoted two-sided errors). 
This yields a best-fit BHFP relation 
\begin{eqnarray}
\label{eqn:fp.sigma.reff}
  \log(\mbh) &=& 8.33 + 3.00(\pm0.30)\,\log(\sigma/200\,{\rm km\,s^{-1}}) \\
\nonumber & & + 0.43(\pm0.19)\,\log(\re/5\,{\rm kpc}) 
\end{eqnarray}
from the observations. 
Unsurprisingly, the slopes in the BHFP relation are close to those formally determined for the 
residuals in Figure~\ref{fig:data.residuals}. 
As expected, the 
residuals of $\mbh$ with respect to these 
fundamental plane relations, at fixed $\re$ and fixed $\sigma$, show no 
systematic trends and are consistent with small intrinsic scatter. The introduction of a 
BHFP eliminates the strong systematic correlations between the residuals, yielding 
flat errors as a function of $\sigma$ and $\re$. 

At low redshift, $\sigma$, $\re$, and $\mdyn$ can be determined reliably, but 
at high redshift it is typically the stellar mass $\mstar$ or luminosity which is used 
to estimate $\mbh$. Therefore, it is interesting to examine the BHFP projections 
in terms of e.g.\ $\mstar$ and $\sigma$ or 
$\mstar$ and $\re$. Repeating our analysis, we 
find in Figure~\ref{fig:data.residuals} 
that the observations demand a FP relation over a simple $\mbh(\mstar)$ relation at 
high significance. The exact values of the best-fit coefficients of this 
BHFP determined from the observations 
are given (along with those of various 
other BHFP projections) in Table~\ref{tbl:correlations}.

\tableclear
\begin{\tableset}{cccc}
\tabletypesize{\scriptsize}
\tablecaption{BH-Host Correlations\label{tbl:correlations}}
\tablewidth{0pt}
\tablehead{
\colhead{Variables\tablenotemark{1}} &
\colhead{Normalization\tablenotemark{2}} &
\colhead{Slope\tablenotemark{3}} &
\colhead{Scatter\tablenotemark{4}} 
}
\startdata
$\sigma^{\alpha}\,\re^{\beta}$ & $8.33\pm0.06$ & $3.00\pm0.30,\ 0.43\pm0.19$ & $0.21$
\\
$\mstar^{\alpha}\,\sigma^{\beta}$ & $8.24\pm0.06$ & $0.54\pm0.17,\ 2.18\pm0.58$ & $0.22$
\\
$\mstar^{\alpha}\,\re^{\beta}$ & $8.06\pm0.07$ & $1.78\pm0.40,\ -1.05\pm0.37$ & $0.25$
\\
$\mstar\,\sigma^{2}$ & $8.23\pm0.06$ & $0.71\pm0.06$ & $0.25$
\\
$\sigma$ & $8.28\pm0.08$ & $3.96\pm0.39$ & $0.31$
\\
$\mstar$ & $8.21\pm0.07$ & $0.98\pm0.10$ & $0.33$
\\
$\mdyn$ & $8.22\pm0.10$ & $1.05\pm0.13$ & $0.43$
\\
$\re$ & $8.44\pm0.10$ & $1.33\pm0.25$ & $0.45$\\
\enddata
\tablenotetext{1}{For the variables $(x,\ y)$, a 
correlation of the form $\log(\mbh)=\alpha\log(x)+\beta\log(y)+\delta$ is assumed, 
where the normalization is $\delta$ and $\alpha$, $\beta$ are the logarithmic slopes.}
\tablenotetext{2}{The normalization gives $\log(\mbh)$ for $\sigma=200\,{\rm km\,s^{-1}}$, 
$\mstar=10^{11}\,M_{\sun}$, $\mdyn=10^{11}\,M_{\sun}$, $\re=5\,{\rm kpc}$, 
which roughly minimizes the covariance between fit parameters.}
\tablenotetext{3}{Errors quoted here for the BHFP relations in $(\sigma,\ \re)$, $(\mstar,\ \sigma)$, and 
$(\mstar, \re)$ include the covariance between the two slopes. Holding one of the two fixed and 
varying the other yields substantially smaller errors (typically $\sim5\%$). All quoted errors 
account for measurement errors in both $\mbh$ and the relevant independent variables.}
\tablenotetext{4}{The internal scatter is estimated from the observations 
as that which yields a reduced $\chi^{2}/\nu=1$ with respect to the given best-fit relation.}
\end{\tableset}
\tableclear

The BHFP in terms of $\sigma$ and $\re$ (i.e.\ $\mbh\propto\sigma^{\alpha}\,\re^{\beta}$) 
is of course tightly related to the BHFP in terms of 
$\sigma$ and $\mstar$ ($\mbh\propto\sigma^{\alpha}\,\mstar^{\beta}$) or 
$\mstar$ and $\re$; the near-IR fundamental plane relates 
stellar mass (assuming $K$-band luminosity is a good proxy for stellar mass), 
effective radius, and velocity dispersion as 
$\re\propto\sigma^{1.53}\,I_{e}^{-0.79}$ \citep{pahre:nir.fp}. 
Using $I_{e}\propto \mstar / \re^{2}$, 
we can substitute this in Equation~(\ref{eqn:fp.sigma.reff}) and obtain the 
expected BHFP in terms of $\sigma$ and $\mstar$, namely 
$\mbh\propto\sigma^{1.9}\,\mstar^{0.58}$. This is quite close to the result of our 
direct fitting, $\mbh\propto\sigma^{2.2}\,\mstar^{0.54}$. 
The FP of early-type galaxies, in other words, allows us to relate our 
two ``fundamental planes,'' namely 
$\mbh\propto \sigma^{3}\,\re^{1/2}$ and $\mbh\propto\sigma^{2}\,\mstar^{1/2}$. 
Given the tight early-type FP relation between $\mstar$, $\sigma$, and $\re$, 
these two forms of the BHFP are completely equivalent (the choice 
between them is purely a matter of convenience). We can, 
in fact, pick any two of the three FP-related variables as our independent 
variables for predicting $\mbh$ -- using the early-type FP again to transform 
the BHFP 
relation in terms of $\sigma$ and $\re$ to one in terms of $\mstar$ and 
$\re$, we expect $\mbh\propto\mstar^{1.6}\,\re^{-0.8}$, 
similar to the relation we directly fit (see Table~\ref{tbl:correlations}). 
Any two of $\mstar$, $\re$, and $\sigma$ can thus be used to predict 
$\mbh$ according to the BHFP relations. The tightness of the early-type FP 
also means that it is redundant to search for a four-variable correlation 
(i.e.\ one of the form $\mbh\propto\sigma^{\alpha}\,\re^{\beta}\,\mstar^{\gamma}$), 
since $\mstar$ is itself a function of $\sigma$ and $\re$ with very 
little scatter (and indeed, directly testing this, we find no significant 
improvement in our fits expanding to a four-variable correlation). 
Perhaps most important, however, is 
that no transformation (given the early-type FP relating these three 
variables) eliminates the dependence on two variables -- i.e.\ 
no transformation from any one BHFP allows us to write 
$\mbh$ as a pure function of either $\sigma$, $\re$, $\mstar$, 
or $\mdyn$ (as we expect, since Figure~\ref{fig:data.residuals} 
explicitly shows that each of these single-variable correlations 
exhibits a significant residual dependence on another variable).

Given the definition of $\mdyn\propto\sigma^{2}\,\re$, it is trivial to convert 
the best-fit  
BHFP relation in terms of $\sigma$ and $\re$ 
to one in $\mdyn$, obtaining
\begin{eqnarray}
  \log(\mbh) &=& 8.29 + 0.43\,\log(\mdyn/10^{11}\,M_{\sun}) \\
\nonumber & & + 2.14\,\log(\sigma/200\,{\rm km\,s^{-1}}) \\ 
\nonumber &=& 8.19 + 1.50\,\log(\mdyn/10^{11}\,M_{\sun}) \\
\nonumber & & - 1.07\,\log(\re/5\,{\rm kpc})  \, .
\end{eqnarray}
It is important to note that the residual correlation with either $\sigma$ or 
$\re$ at fixed $\mdyn$ is non-zero and highly significant. It is therefore not the 
case that the BHFP reflects, for example, the true correlation being 
between $\mbh$ and $\mdyn$ (in which case the BHFP would have a form 
$\mbh\propto (\sigma^{2}\,\re)^{\gamma}$), but it is a FP in a genuine sense. 

We have also repeated the analysis of Figure~\ref{fig:data.residuals} 
for $\mdyn$ and $\sigma$ (or $\mdyn$ and $\re$), and obtain these results directly, 
with a pure correlation between $\mbh$ and $\mdyn$ ruled out 
at $\sim3\sigma$ in the observations. We note 
that any analysis of these particular residual correlations must 
be done with special care, as the plotted 
quantities (e.g.\ $\mbh/\mdyn$ versus $\re / \langle\re[\mdyn]\rangle$) are {\em not} 
independent (since $\mdyn \propto \sigma^{2}\,\re/G$ depends directly on the measured $\re$), 
and so fitting a small sample where there is both intrinsic scatter and measurement errors 
in the quantities can bias the fit. Running a series of Monte Carlo experiments allowing for 
the range of estimated intrinsic scatter and measurement errors in each quantity, we 
find that the observations span a sufficiently small baseline that, with the present 
errors, a naive comparison will be biased to {\em underestimate} the significance of the 
preference for a BHFP relation over a simple correlation with $\mdyn$. 
The $P_{\rm null}$ for correlations between the residuals in 
the $\mbh-\mdyn$ relation and $\re$ or $\sigma$ typically decreases by a factor $\sim2$ 
with a proper Monte Carlo analysis; i.e.\ the 
true significance of the BHFP relation is even greater than a direct comparison 
suggests. 

\begin{figure}
    \centering
    \figexpand
    \plotter{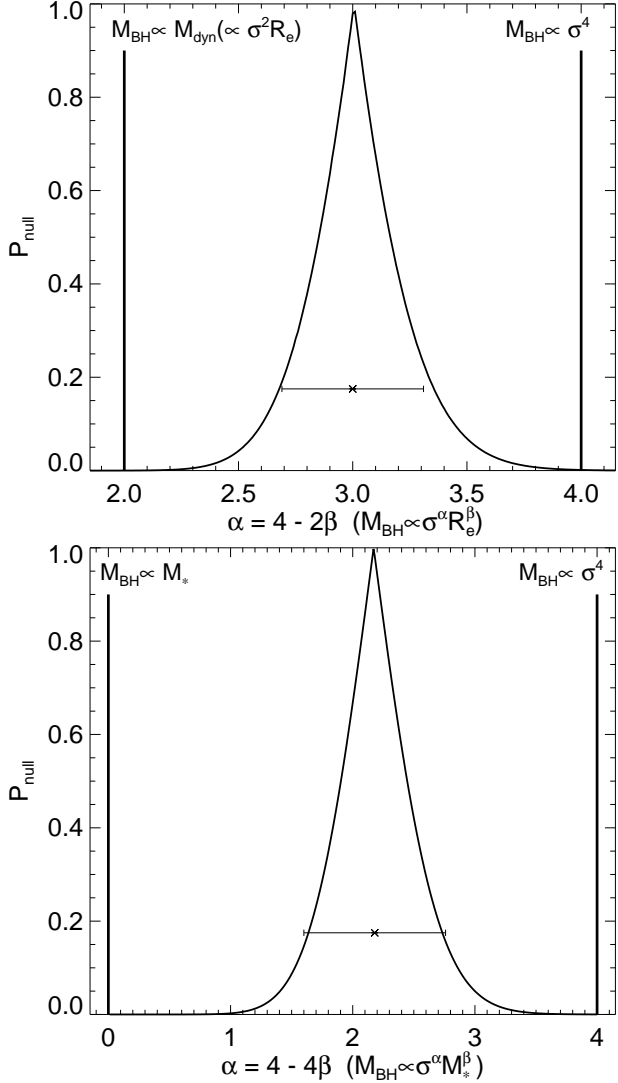}
    \caption{Probability that there is no remaining trend or correlation among 
    the residuals of the $M_{\rm BH}$-host relation ($P_{\rm null}$, as in 
    Figure~\ref{fig:data.residuals}), 
    as a function of the slope $\alpha$ for a correlation of the form 
    $M_{\rm BH}\propto \sigma^{\alpha}\,\re^{\beta}$ ({\em top}) or 
    $M_{\rm BH}\propto \sigma^{\alpha}\,\mstar^{\beta}$ ({\em bottom}). 
    For each value of $\alpha$, we marginalize over $\beta$ and the 
    correlation normalization to find the best fit, but there 
    is a rough degeneracy between the best fit $\alpha$ and $\beta$ 
    ($\beta\approx (4-\alpha)/2$ for $\mbh\propto\sigma^{\alpha}\,\re^{\beta}$, 
    $\beta\approx (4-\alpha)/4$ for $\mbh\propto\sigma^{\alpha}\,\mstar^{\beta}$). 
    Lines show $P_{\rm null}$ for the observations, points show the 
    best fit (in a $\chi^{2}$ sense) BHFP relation and errors 
    from Table~\ref{tbl:correlations}. Thick black lines show the 
    $\alpha$ corresponding to a pure $\mbh-\sigma$, $\mbh-\mdyn$, or 
    $\mbh-\mstar$ relation, all of which are ruled out at similar 
    ($\sim3\,\sigma$) significance. 
    \label{fig:degeneracy}}
\end{figure}

Figure~\ref{fig:data.residuals} demonstrates the significance with 
which the observations rule out {\em both} a 
pure BH-host mass relation (either $\mbh\propto\mstar$ or 
$\mbh\propto\mdyn$) and a pure $\mbh-\sigma$ relation. However, 
when fitting to a form $\mbh\propto\sigma^{\alpha}\,\re^{\beta}$, 
there is still some degeneracy between the slopes $\alpha$ and 
$\beta$ (roughly along the axis $\beta\approx(4-\alpha)/2$).
Figure~\ref{fig:degeneracy} 
illustrates the degree of this degeneracy and the extent to which, for example, 
a BHFP with $\mbh\propto \sigma^{3}\,\re^{1/2}$ is favored over a pure 
$\mbh-\mdyn$ relation. We plot the 
likelihood of a residual correlation between $\mbh$ and $\re$ or $\sigma$ 
at fixed $\sigma^{\alpha}\,\re^{\beta}$, as a function of the 
slope $\alpha$ (marginalizing over $\beta$ and other fit parameters 
at each $\alpha$, although $\beta(\alpha)$ roughly follows the axis of 
degeneracy above). 
In detail, this is the 
likelihood of a correlation between the 
residuals of the $\mbh\propto\sigma^{\alpha}\,\re^{\beta}$ relation and 
the residuals of the $\sigma(\sigma^{\alpha}\,\re^{\beta})$ and 
$\re(\sigma^{\alpha}\,\re^{\beta})$ relations, identical to the 
procedure used to determine the quoted $P_{\rm null}$ in 
Figure~\ref{fig:data.residuals}. 
We compare the quoted best-fit slopes from Table~\ref{tbl:correlations}. 
We then repeat this exercise for the alternate representation of the 
fundamental plane, $\mbh\propto\sigma^{\alpha}\,\mstar^{\beta}$. 

The analysis shown in Figure~\ref{fig:degeneracy} agrees reasonably
well with the $\chi^{2}$ best-fit expectations, and illustrate an
important point: a pure relation between BH and host mass (either
dynamical mass $\mdyn$ or stellar mass $\mstar$) is ruled out at a
significance level comparable to that with which a pure $\mbh-\sigma$
relation is ruled out ($\sim3\,\sigma$). Indeed, the preferred BHFP
parameters are centered almost exactly mid-way between these two
previously proposed relations.

However, there are possible correlations which cannot be clearly
discriminated by the present data.  Both a pure relation between
BH mass and spheroid binding energy, of the form
$\mbh\propto(\mstar\,\sigma^{2})^{2/3}$, as studied in
\citet{aller:mbh.esph}, for example, and a mixed relation of the
form $\mbh\propto\mstar^{1/2}\,\sigma^{2}$
\citep[see][]{hopkins:bhfp.theory}, are within the $\sim1\sigma$
allowed range of the data.  It is worth noting that simply
expanding the number of observed sources will not necessarily break
these degeneracies. Rather, to increase the constraining power of the
observations, a larger baseline is needed, including (in particular) a
larger sample of objects which lie off the mean $\re-\sigma$ or
$\mstar-\sigma$ relations (and thus extend the baseline in the
residual-residual space which properly constrains the FP slopes).

Finally, we noted above that Figure~\ref{fig:data.residuals} 
is essentially unchanged if we consider residuals with respect to 
just linear (i.e.\ pure power law) $\mbh-\sigma$-type relations, as a 
consequence of there being no significant evidence in our data for a 
log-quadratic or higher-order correlation. Allowing log-quadratic terms in 
our fundamental plane fits, we find a best fit to the observations of
the form
\begin{eqnarray}
\log(\mbh) & =8.06& + (2.8\pm0.4)\,\tilde{\sigma} + (0.48\pm0.18)\,\tilde{r} \\
\nonumber & & -  (2.1\pm2.3)\,\tilde{\sigma}^{2} + (0.31\pm 0.25)\,\tilde{r}^{2}
\end{eqnarray}
(where $\tilde{\sigma}\equiv\log{[\sigma/{\rm 200\,km\,s^{-1}}]}$ and 
$\tilde{r}\equiv\log{[\re/{\rm 3\,kpc}]}$). 
The linear BHFP coefficients in $\sigma$ and $\re$ are similar to those in 
Table~\ref{tbl:correlations}, and their significance is not much changed -- 
this illustrates that the FP behavior we find cannot simply trade off with or 
be equally well-represented 
by a log-quadratic dependence (i.e.\ one cannot eliminate the 
residual dependence of $\mbh$ on $\re$ at fixed $\sigma$ by adding a 
log-quadratic or higher-order term in $\sigma$). 
The log-quadratic terms are at most significant at the 
$\sim1\sigma$ level. This is also 
true if we add just one of the two log-quadratic terms -- 
adding a log-quadratic term in just $\re$ 
yields a coefficient $(0.16\pm0.25)\,\tilde{r}^{2}$, and adding one in just $\sigma$ gives a 
coefficient $(-0.28\pm2.15)\,\tilde{\sigma}^{2}$. This is similar to the finding
of \citet{wyithe:log.quadratic.msigma}, who estimates $<1\sigma$ significance for 
the addition of log-quadratic 
terms in any of $\sigma$, $\mdyn$, or $\mstar$.

\section{Conclusions}
\label{sec:conclusions}

We study the correlation between observed central BH mass and host
galaxy properties, and find that the systems lie on a BH ``fundamental
plane'' (BHFP), of the form $\mbh\propto\sigma^{3.0}\,\re^{0.5}$ or
$\mbh\propto \mstar^{0.5-0.7}\,\sigma^{1.5-2.0}$, analogous to the FP
of spheroids.  Specifically, there are significant (at $>99.9\%$
confidence) trends in the residuals of the $\mbh-\sigma$ relation with
$\mstar$ and $\re$ at fixed $\sigma$, and likewise in the
$\mbh-\mstar$ relation (with $\sigma$ or $\re$ at fixed $\mstar$).
This provides a new paradigm for understanding the traditional
relations between BH mass and either bulge velocity dispersion or
mass. These correlations (as well as those with other bulge properties
such as effective radius, central potential, dynamical mass,
concentration, Sersic index, and bulge binding energy) are all
projections of the same fundamental plane relation.  Just as the
Faber-Jackson relation between e.g.\ stellar mass or luminosity and
velocity dispersion ($\mstar-\sigma$) is understood as a projection of
the more fundamental relation between $\mstar$, $\sigma$, and $\re$,
so too is the $\mbh-\sigma$ relation ($\mbh\propto\sigma^{4}$) a
projection of the more fundamental relation $\mbh\propto
\sigma^{3}\,\re^{0.5}$. Recognizing this resolves the nature of
several apparent outliers in the $\mbh-\sigma$ relation, which
simply have unusual $\sigma$ values for their stellar masses or
effective radii, and eliminates the strong correlations between
residuals.

Improved measurements of the host properties of systems with 
well-measured BHs can significantly improve constraints on the BHFP. As noted 
in Table~\ref{tbl:correlations}, the present observations demand a correlation of the 
form $\mbh\propto\sigma^{\alpha}\,\mstar^{\beta}$ over a simple correlation with 
either $\sigma$ or $\mstar$ at $\gtrsim3\,\sigma$ confidence. Already, 
this puts strong constraints on theoretical models of BH growth and evolution -- 
BH mass does not simply scale with the star formation (stellar mass) 
or virial velocity of the host galaxy. 
However, there is still a 
substantial degeneracy between the slopes $\alpha$ and $\beta$ (roughly 
along the axis $\beta\approx1-\alpha/4$). For example, the existing data do not 
allow us to significantly distinguish a pure correlation with spheroid binding 
energy $\mbh\propto (\mstar\,\sigma^{2})^{2/3}$, as detailed in 
\citet{aller:mbh.esph} from the marginally favored 
relation $\propto \mstar^{1/2}\,\sigma^{2}$.  Both suggest 
that the ability of BHs to self-regulate 
their growth must be sensitive to the potential well at the 
center of the galaxy (and therefore to galactic structure), but the difference 
could reveal variations in the means by which BH feedback couples 
to the gas on these scales. 

Increasing the observed sample sizes and, 
in particular, extending the observed baselines in mass and $\sigma$ will 
substantially improve the lever arm on these correlations. In particular, the addition 
of stellar mass $\mstar$ information to the significant number of objects which have 
measurements of $\sigma$ and indirect measurements of $\mbh$ from 
reverberation mapping would enable considerably stronger tests of our 
proposed BHFP relation. 
We do note the caveat, however, 
that care should 
still be taken to consider only bulge properties and remove e.g.\ rotationally supported 
contributions to the velocity dispersion.

The BHFP appears to be a robust correlation, which provides an improved 
context in which to understand the nature and evolution of the numerous observed 
correlations between BH and host spheroid properties.  In particular, the results
described here provide new, important 
constraints for models of BH growth, feedback and self-regulation.

\acknowledgments We thank Chien Peng for illuminating discussion and comments. 
We also thank an anonymous referee for suggestions that greatly
improved the clarity of the manuscript.
This work was supported in part by NSF grant AST
03-07690, and NASA ATP grants NAG5-12140, NAG5-13292, and NAG5-13381.

\begin{appendix}
\section{Determining the Significance of Residuals in the $\mbh$-Host Correlations}
\label{sec:appendix}

We test the robustness of our results, and how appropriate our fitting method is, by 
constructing a series of Monte Carlo realizations of the observations. 
As the velocity distribution $\sigma$ is the most well-measured quantity for 
the observed galaxies (with the exception of the Milky Way, which we generally 
exclude from our fits following \citet{tremaine:msigma}), we begin with the observed distribution 
of $\sigma$ values. Either from the cumulative distribution of $\sigma$ values, or 
from the observed error bars for each value of $\sigma$, we then 
statistically resample the $\sigma$ distribution (with the same number of objects) 
for each Monte Carlo realization. Next, we assume a mean intrinsic correlation 
between $\re$ and $\sigma$ (for simplicity, assume $\re\propto\sigma^{2}$, 
similar to that observed), with some intrinsic scatter comparable to 
that observationally inferred, and use this to randomly generate the true 
$\re$ values of each point. Then, given some assumed true BH-host correlation 
of the form $\mbh\propto\sigma^{\alpha}\,\re^{\beta}$, again with an 
intrinsic scatter comparable to the observational estimates, we randomly 
generate the true BH masses. Finally, using the mean observational 
measurement errors in each of 
these quantities (or, specifically the quoted observational errors of each object 
corresponding to each mock point, as it makes little difference), we randomly 
generate the ``observed'' values of each quantity. We then repeat our fitting 
procedures from \S~\ref{sec:local:FP}. For example, to search for residuals with respect 
to the $\mbh-\sigma$ relation, we fit the ``observed'' points to a mean 
$\mbh-\sigma$ relation, and compare the residuals in BH mass to 
the residuals in $\re$ of the ``observed'' $\re-\sigma$ relation. We determine the 
statistical significance of these residuals, and fit to determine the 
residual dependence (i.e.\ best fit ``observed'' $\mbh\propto\sigma^{\alpha}\,\re^{\beta}$ 
relation). 

\begin{figure*}
    \centering
    \plotone{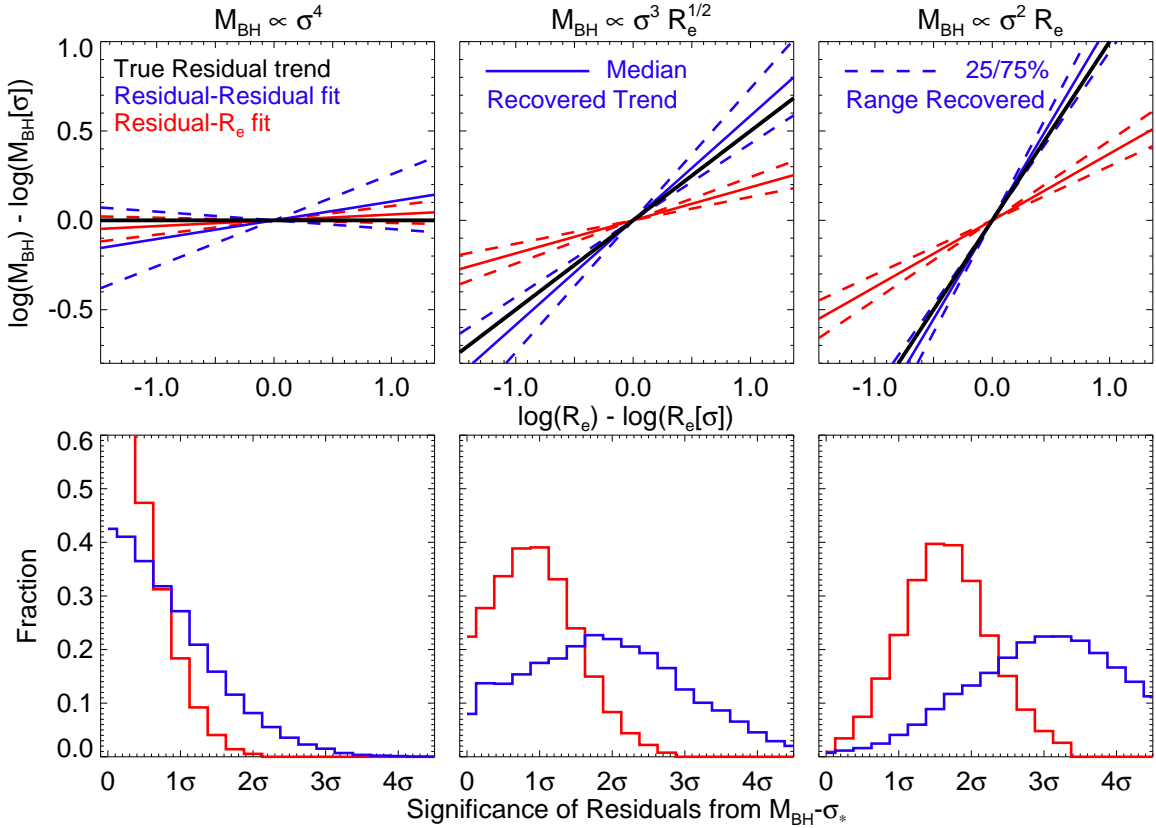}
    \caption{{\em Top:} Recovered dependence of $\mbh$ on $\re$ 
    at fixed $\sigma$ (blue lines) from Monte Carlo realizations of the 
    observed data sets, assuming different ``true'' intrinsic correlations 
    between $\mbh$ and host properties 
    (of the form $\mbh\propto\sigma^{\alpha}\,\re^{\beta}$). 
    Black lines in each panel show the residual of the assumed ``true'' 
    relation, i.e.\ zero dependence in the case where the 
    true relation is $\mbh\propto\sigma^{4}$ ({\em left}), 
    residual $\mbh\propto\re^{1/2}$ for a BHFP relation 
    of the form $\mbh\propto\sigma^{3}\,\re^{1/2}$ ({\em center}), and 
    residual $\mbh\propto\re$ for a $\mbh\propto\mdyn$ relation ({\em right}). 
    Solid blue lines show the median dependence recovered using our 
    standard fitting method (fitting residuals at fixed $\sigma$) 
    for all Monte Carlo realizations, dashed lines show the $25/75\%$ 
    range of fits recovered. If we fit the residuals as a function of 
    the actual value of $\re$ (instead of $\re$ at fixed $\sigma$), 
    we infer the correlations shown as red lines. In all cases, 
    fitting residuals versus residuals (our standard methodology) 
    recovers a similar slope to the ``true'' correlation and appropriate 
    (in a $\chi^{2}$ sense for normal errors) range about that true slope. However, 
    comparing to the actual value of $\re$ severely smears out 
    the true dependence on $\re$ when it is present. 
    {\em Bottom:} The distribution of the significance of the residuals (of 
    $\mbh$ with respect to $\re$) in our Monte Carlo realizations. 
    The same trends are seen: fitting residuals versus residuals yields 
    the expected $\chi^{2}$ distributions and generally detects the appropriate 
    significance of residuals when there is a true dependence on $\re$, 
    but in almost all cases (even when the true correlation has a 
    strong dependence on $\re$), fitting residuals in $\mbh$ against the 
    actual value of $\re$ leads one to conclude (incorrectly) that 
    the residuals are not significant.    
    \label{fig:demo.fits}}
\end{figure*}

Figure~\ref{fig:demo.fits} shows the results of this analysis. Specifically, we show 
the median correlations between the residuals in $\mbh$ and in $\re$, from 
$\sim1000$ Monte Carlo realizations, and the $25/75\%$ quartile ranges of 
the fitted correlations. We show this for three assumed cases -- one in which 
the ``true'' correlation is a pure $\mbh-\sigma$ relation ($\mbh\propto\sigma^{4}$), 
one in which the true correlation is similar to our best-fit BHFP 
relation ($\mbh\propto\sigma^{3}\,\re^{1/2}$), and one in which the true 
correlation is a pure $\mbh-\mdyn$ relation ($\mbh\propto\mdyn\propto\sigma^{2}\,\re$). 
In all of these cases, our fitting method generally recovers a residual 
correlation very similar to the input correlation. There appears to be a 
(very slight) bias towards our fitting method recovering a marginally stronger 
dependence of $\mbh$ on $\re$, however, in terms of its statistical significance, 
the bias introduced is very small -- at most, it implies that the ``true'' (in 
a maximum likelihood sense) significance of the dependence of $\mbh$ on 
$\re$ at fixed $\sigma$ which we recover from the observations should be 
slightly lower than quoted (equivalent to increasing our error bars by $\sim10\%$). 
In other words, the 
observational data span a sufficient baseline, and the fitting method we use 
is sufficiently robust, that any ``true'' correlation of the form 
$\mbh\propto\sigma^{\alpha}\,\re^{\beta}$ should be recovered in 
the majority of cases. 

We can also check the significance of the residuals (of $\mbh$ with respect to 
$\re$) from each of these Monte Carlo realizations, as shown in 
Figure~\ref{fig:demo.fits}. In the case of a pure ``true'' $\mbh-\sigma$ relation, 
the residuals should not be significant in most cases, and indeed we find 
that they are not. Their behavior is essentially exactly what is expected for 
a $\chi^{2}$ distribution -- i.e.\ in $5\%$ of all cases, a $95\%$ significance 
level is assigned to the residuals, in $1\%$ of all cases, a $99\%$ 
significance level is assigned. The significance as we have determined it 
therefore carries its appropriate (expected) meaning and weight, 
and our general $\chi^{2}$ analysis and assumption of normal errors 
in log-space is probably not misleading. In the case where the 
``true'' correlation is of the form $\mbh\propto\sigma^{3}\,\re^{1/2}$ or 
$\mbh\propto\sigma^{2}\,\re$, the likelihood that the 
residuals will be significant increases dramatically (as it should), 
although we caution that there is still some (fairly large) probability 
that the residuals of a given Monte Carlo realization will not be 
especially significant ($\lesssim2\,\sigma$ significance). From 
comparison with the residuals with respect to a pure $\mbh-\sigma$ 
relation, this implies that the parameter space spanned by current 
observations is such that a significant detection of residuals 
{\em does} imply a significant probability of a true BHFP relation, 
but that a weak or non-detection of such residuals 
{\em does not yet rule out} such a relation. 

We caution, however, that this is only the case when comparing residuals 
against residuals. If we were, as in Figure~\ref{fig:demo.residuals}, 
to simply consider the residual in $\mbh-\sigma$ as a function of the 
actual value of $\re$ (instead of comparing $\mbh$ and $\re$ both 
at fixed $\sigma$), we smear out the significance of 
any real residuals, as we would expect. The slope recovered 
(i.e.\ the inferred dependence of $\mbh$ on $\re$) 
is severely biased towards being too shallow for any 
non-zero dependence on $\re$, and in only $\sim1\%$ of cases will 
such a method 
recover a slope similar to the ``true'' intrinsic correlation. 
Looking at the significance of the residuals in this space, 
it is clear that this projection biases against detecting any significant 
residual dependence on $\re$. Even if the true relation 
were $\mbh\propto\sigma^{2}\,\re$, looking at the significance of the 
residuals as a function of the actual value of $\re$ 
would lead one to conclude in over 
$80\%$ of cases that they were insignificant ($<2\,\sigma$ 
formal significance). A proper analysis, however, 
would recover the true significance of the residuals 
in $\gtrsim95-99\%$ of cases. 

Finally, by repeating this analysis over the 
entire parameter space of ``true'' correlations (for example forms 
$\mbh\propto\sigma^{\alpha}\,\re^{\beta}$), 
and comparing with the data, 
we can use a similar Monte Carlo approach to determine a 
maximum-likelihood, best-fit BHFP relation. We have done 
so, and find nearly identical results to our standard fits, again 
indicating that the conclusions herein are robust to the exact 
likelihood calculation so long as the residuals are properly 
analyzed. Similarly, we can repeat this entire analysis 
for true correlations of the 
form $\mbh\propto\sigma^{\alpha}\,\mstar^{\beta}$ or
$\mbh\propto\mstar^{\alpha}\,\re^{\beta}$, or 
in order to test the significance of residuals at fixed $\re$ 
or fixed $\mstar$. In all cases, we recover very similar results, 
reinforcing the robustness of our conclusions.

\end{appendix}

\bibliography{ms}

\begin{thebibliography}{31}
\expandafter\ifx\csname natexlab\endcsname\relax\def\natexlab#1{#1}\fi

\bibitem[{{Aller} \& {Richstone}(2007)}]{aller:mbh.esph}
{Aller}, M.~C., \& {Richstone}, D.~O. 2007, \apj, accepted, arXiv:0705.1165v1
  [astro-ph], 705

\bibitem[{{Batcheldor} {et~al.}(2006){Batcheldor}, {Marconi}, {Merritt}, \&
  {Axon}}]{batcheldor:bcgs}
{Batcheldor}, D., {Marconi}, A., {Merritt}, D., \& {Axon}, D.~J. 2006, \apjl,
  in press [astro-ph/0610264]

\bibitem[{{Bell} {et~al.}(2003){Bell}, {McIntosh}, {Katz}, \&
  {Weinberg}}]{bell:mfs}
{Bell}, E.~F., {McIntosh}, D.~H., {Katz}, N., \& {Weinberg}, M.~D. 2003, \apjs,
  149, 289

\bibitem[{{Bernardi} {et~al.}(2006){Bernardi}, {Sheth}, {Tundo}, \&
  {Hyde}}]{bernardi:magorrian.bias}
{Bernardi}, M., {Sheth}, R.~K., {Tundo}, E., \& {Hyde}, J.~B. 2006, \apj, in
  press [astro-ph/0609300]

\bibitem[{{Bernardi} {et~al.}(2003{\natexlab{a}})}]{bernardi:correlations}
{Bernardi}, M., {et~al.} 2003{\natexlab{a}}, \apj, 125, 1849

\bibitem[{{Bernardi} {et~al.}(2003{\natexlab{b}})}]{bernardi:fp}
---. 2003{\natexlab{b}}, \aj, 125, 1866

\bibitem[{{de Francesco} {et~al.}(2006){de Francesco}, {Capetti}, \&
  {Marconi}}]{defrancesco:mbh.mdyn}
{de Francesco}, G., {Capetti}, A., \& {Marconi}, A. 2006, \aap, 460, 439

\bibitem[{{Djorgovski} \& {Davis}(1987)}]{dd87:fp}
{Djorgovski}, S., \& {Davis}, M. 1987, \apj, 313, 59

\bibitem[{{Dressler} {et~al.}(1987){Dressler}, {Lynden-Bell}, {Burstein},
  {Davies}, {Faber}, {Terlevich}, \& {Wegner}}]{dressler87:fp}
{Dressler}, A., {Lynden-Bell}, D., {Burstein}, D., {Davies}, R.~L., {Faber},
  S.~M., {Terlevich}, R., \& {Wegner}, G. 1987, \apj, 313, 42

\bibitem[{{Faber} \& {Jackson}(1976)}]{fj76}
{Faber}, S.~M., \& {Jackson}, R.~E. 1976, \apj, 204, 668

\bibitem[{{Ferrarese} \& {Merritt}(2000)}]{FM00}
{Ferrarese}, L., \& {Merritt}, D. 2000, \apjl, 539, L9

\bibitem[{{Gebhardt} {et~al.}(2000)}]{Gebhardt00}
{Gebhardt}, K., {et~al.} 2000, \apjl, 539, L13

\bibitem[{{Graham} \& {Driver}(2006)}]{graham:sersic}
{Graham}, A.~W., \& {Driver}, S.~P. 2006, \apj, in press [astro-ph/0607378]

\bibitem[{{Graham} {et~al.}(2001){Graham}, {Erwin}, {Caon}, \&
  {Trujillo}}]{graham:concentration}
{Graham}, A.~W., {Erwin}, P., {Caon}, N., \& {Trujillo}, I. 2001, \apjl, 563,
  L11

\bibitem[{{H{\"a}ring} \& {Rix}(2004)}]{haringrix}
{H{\"a}ring}, N., \& {Rix}, H.-W. 2004, \apjl, 604, L89

\bibitem[{{Hopkins} {et~al.}(2007){Hopkins}, {Hernquist}, {Cox}, {Robertson},
  \& {Krause}}]{hopkins:bhfp.theory}
{Hopkins}, P.~F., {Hernquist}, L., {Cox}, T.~J., {Robertson}, B., \& {Krause},
  E. 2007, \apj, in press [astro-ph/0701351]

\bibitem[{{Kormendy}(1977)}]{kormendy77:correlations}
{Kormendy}, J. 1977, \apj, 218, 333

\bibitem[{{Kormendy} {et~al.}(2007){Kormendy}, {Fisher}, {Cornell}, \&
  {Bender}}]{kormendy:wetvsdry}
{Kormendy}, J., {Fisher}, D.~B., {Cornell}, M.~E., \& {Bender}, R. 2007, \apj,
  submitted

\bibitem[{{Kormendy} \& {Richstone}(1995)}]{KormendyRichstone95}
{Kormendy}, J., \& {Richstone}, D. 1995, \araa, 33, 581

\bibitem[{{Lauer} {et~al.}(2005)}]{lauer:centers}
{Lauer}, T.~R., {et~al.} 2005, \aj, 129, 2138

\bibitem[{{Lauer} {et~al.}(2006{\natexlab{a}})}]{lauer:bimodal.profiles}
---. 2006{\natexlab{a}}, \apj, in press [astro-ph/0609762]

\bibitem[{{Lauer} {et~al.}(2006{\natexlab{b}})}]{lauer:massive.bhs}
---. 2006{\natexlab{b}}, \apj, in press [astro-ph/0606739]

\bibitem[{{Magorrian} {et~al.}(1998)}]{magorrian}
{Magorrian}, J., {et~al.} 1998, \aj, 115, 2285

\bibitem[{{Marconi} \& {Hunt}(2003)}]{marconihunt}
{Marconi}, A., \& {Hunt}, L.~K. 2003, \apjl, 589, L21

\bibitem[{{McDermid} {et~al.}(2006)}]{mcdermid:sauron.profiles}
{McDermid}, R.~M., {et~al.} 2006, \mnras, 1312

\bibitem[{{Merritt} \& {Ferrarese}(2001)}]{merrittferrarese:msigma}
{Merritt}, D., \& {Ferrarese}, L. 2001, \apj, 547, 140

\bibitem[{{Novak} {et~al.}(2006){Novak}, {Faber}, \& {Dekel}}]{novak:scatter}
{Novak}, G.~S., {Faber}, S.~M., \& {Dekel}, A. 2006, \apj, 637, 96

\bibitem[{{Pahre} {et~al.}(1998){Pahre}, {Djorgovski}, \& {de
  Carvalho}}]{pahre:nir.fp}
{Pahre}, M.~A., {Djorgovski}, S.~G., \& {de Carvalho}, R.~R. 1998, \aj, 116,
  1591

\bibitem[{{Shen} {et~al.}(2003){Shen}, {Mo}, {White}, {Blanton}, {Kauffmann},
  {Voges}, {Brinkmann}, \& {Csabai}}]{shen:size.mass}
{Shen}, S., {Mo}, H.~J., {White}, S.~D.~M., {Blanton}, M.~R., {Kauffmann}, G.,
  {Voges}, W., {Brinkmann}, J., \& {Csabai}, I. 2003, \mnras, 343, 978

\bibitem[{{Tremaine} {et~al.}(2002)}]{tremaine:msigma}
{Tremaine}, S., {et~al.} 2002, \apj, 574, 740

\bibitem[{{Wyithe}(2006)}]{wyithe:log.quadratic.msigma}
{Wyithe}, J.~S.~B. 2006, \mnras, 365, 1082

\end{thebibliography}

\end{document}